\documentclass[reprint,aps,prb,floatfix,superscriptaddress]{revtex4-2}
\usepackage{amsmath,amssymb}
\usepackage{graphicx}
\usepackage[dvipsnames]{xcolor}
\usepackage{dsfont}

\newcommand{\bvec}{\mathbf}

\begin{document}
\title{Ultrafast electron-phonon scattering in an antiferromagnetic Dirac semimetal}

\author{Marius Weber}
\author{Kai Leckron}
    \affiliation{Department of Physics and Research Center OPTIMAS, University of Kaiserslautern-Landau, 67663 Kaiserslautern, Germany}
\author{Libor \v{S}mejkal}
\author{Jairo Sinova}
    \affiliation{Institut f\"{u}r Physik, Johannes Gutenberg University Mainz, 55099 Mainz, Germany}
\author{Baerbel Rethfeld}
\author{Hans Christian Schneider}
    \affiliation{Department of Physics and Research Center OPTIMAS, University of Kaiserslautern-Landau, 67663 Kaiserslautern, Germany}

\date{\today}

\begin{abstract}
Topological antiferromagnetic systems, which exhibit anisotropic band structures combined with complex relativistic spin structures in momentum space, have shown strong magnetoresistance effects driven by Dirac fermion characteristics. While these new antiferromagnets have been studied in transport experiments, 
very little is known about their spin-dependent electronic dynamics on ultrafast timescales and far-from-equilibrium behavior. This paper investigates theoretically the spin-dependent electronic dynamics due to electron-phonon scattering in a model electronic band structure that corresponds to a Dirac semimetal antiferromagnet. Following a spin conserving instantaneous excitation we obtain a change of the antiferromagnetic spin polarization due to the scattering dynamics for the site-resolved spin expectation values. This allows us to identify fingerprints of the anisotropic band structure in the carrier dynamics on ultrashort timescales.
\end{abstract}


\maketitle

\section{introduction}
The field of antiferromagnetism has seen a surge in interest over the last years due to its potential for applications in current-based spintronics~\cite{Jungwirth2016,Wadley,Smejkal2018,Nemec2018,Tserkovnyak2018}. One particularly appealing property of the antiferromagnetic (AFM) phase is its robustness against magnetic perturbations. Further, the characteristics of magnonic excitations in antiferromagnets are very different from ferromagnets and their resonance frequencies make it possible to address them directly by THz fields. However, their insensitivity to magnetic fields also leads to a big challenge, as it is more difficult to excite and measure  their order parameter, i.e., the N\'eel vector. This sets them apart from ferromagnetic materials, where the magnetization may be manipulated relatively easily by external fields, but which suffer, among other things, from stray fields and thermal stability. 

\begin{figure}[b!]
    \centering
    \includegraphics[width=\linewidth]{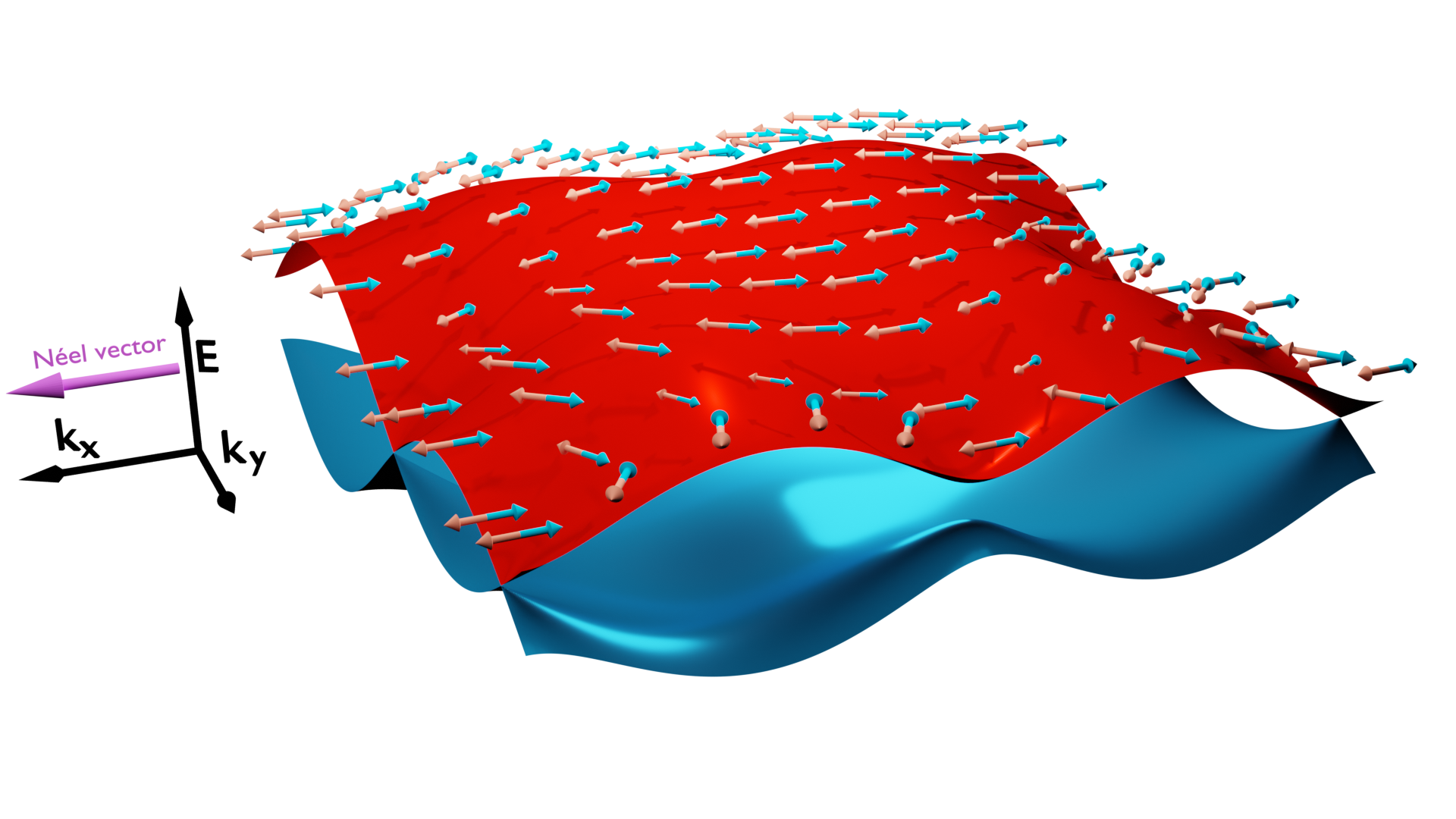}
    \caption{Band structure of the minimal model of an antiferromagnetic Dirac-semimetal. The arrows on top of the upper two degenerate bands (red surface) represent the $k$-local spin expectation values for each band, blue arrows correspond to the upper `$-$'-band and red arrows to the upper `$+$'-band described in the main text.}
    \label{fig:band-structure}
\end{figure}

Only recently, there have been attempts to study the excitation of AFMs by optical fields on ultrashort timescales~\cite{rongione_natcomm_2023,wust_indirect_unpublished_2022,Meer-AFM}. While the coupling of the spin-dependent electronic dynamics to the AFM order parameter is currently being investigated~\cite{ono_npj_2021}, it is also of interest to understand the dynamics of photoexcited carriers for which the microscopic details of the electronic scattering processes might be important.

This paper presents a theoretical study of the spin-dependent electronic dynamics in an AFM model system due to incoherent scattering processes using a microscopic approach. We employ the minimal AFM model recently proposed by \v{S}mejkal et al.~\cite{smejkal_route_2017}. It consists of two doubly degenerate bands and exhibits a band structure with pronounced anisotropies including two Dirac points. For this system, we compute the microscopic electron-phonon scattering dynamics, which leads, in conjunction with spin-orbit coupling in the electronic states, to spin flip/spin relaxation processes. This type of spin-dependent dynamics of electrons in degenerate bands originating from a scattering mechanism that by itself is spin-independent (or very weakly spin dependent) is the Elliott-Yafet mechanism. While the Elliott-Yafet  mechanism was originally proposed for electron-spin resonance~\cite{Yafet1963}, \emph{dynamical} calculations of momentum resolved occupation numbers~\cite{baral_magnetic_2015,essert_electron-phonon_2011,koopmans_explaining_2010,wu_spin_2010,mueller_feedback_2013} can also describe far from equilibrium effects as they arise after optical excitation with ultrashort pulses. However, such microscopic momentum resolved calculations are numerically demanding so that calculations have often been limited to simplified band structures consisting of only a few bands and/or using isotropic band structures~\cite{krauss_2009,leckron_ultrafast_2017}. Effects due to the anisotropic character of the band structure may be captured, for instance, by limiting the $\bvec{k}$-space to an irreducible wedge and a low number of $\bvec{k}$-points~\cite{essert_electron-phonon_2011}, or projecting energy-dependent occupation numbers back to momenta through a band structure mapping~\cite{mueller_relaxation_2013, caruso}. Our first goal is to explore electronic excitations in the spin and energy ``landscape'' presented by topological antiferromagnets as introduced in~\cite{smejkal_electric_2017} with a dynamical approach that is free from the above-mentioned restrictions. This allows us to determine the consequences of incoherent electron-phonon scattering in the full Brillouin-zone of the AFM without approximations that result from placing symmetry or isotropy restrictions on the dynamical distribution functions. We will, however, not attempt a study of the interplay of optical fields with elementary excitations at low densities, which has been shown to lead to cascade-like dynamics in graphene~\cite{vasko-interband-saturation}.

We stress that the anisotropic spin structure has two facets in antiferromagnets: First, in any Kramers degenerate band structure one has opposite expectation values of the (vector) spin in the degnerate Bloch states at each $\bvec{k}$. Second, the antiferromagnetic order is reflected in the different spins in the magnetic unit cell, for instance in an A-B structure in real space, as in the present model. Investigating the dynamics of both spin structures on ultrafast timescales is a central goal of this paper.

This paper is organized as follows. In Sec.~2 we present the model band structure and discuss our parameters and the resulting spin texture. Further, we introduce the equation of motion for the electronic distribution function due to electron-phonon scattering. In Sec.~3 we present the calculated spin-dependent electron dynamics after an instantaneous excitation process. We discuss the fingerprints of the band structure in the electronic dynamics and the consequences of electronic scattering processes for the antiferromagnetic spin polarization. We present our conclusions in Sec.~4. 

\section{Theory}

We first discuss the model used to describe the electronic states of a prototypical topological antiferromagnet. We employ the minimal model introduced in Refs.~\cite{smejkal_electric_2017, smejkal_route_2017} to study the anomalous transport properties of these systems. It is defined for a two-dimensional $\bvec{k} = (k_x,k_y)$-space by the effective Hamiltonian
\begin{equation}
    \begin{split}
    H(\bvec{k}) & =A(\bvec{k})\tau_x  \otimes \mathds{1}+\alpha(\bvec{k})\tau_z \otimes \sigma_x +\beta(\bvec{k}) \tau_z \otimes \sigma_y\\ & \quad \mbox{}+ J_{z} \tau_z \otimes \sigma_z,
    \end{split}
    \label{eq:hamiltonian}
\end{equation}
which is expressed in terms of the momentum-dependent coefficients $A(\bvec{k})= -2t \, \cos(\frac{k_x}{2})\cos(\frac{k_y}{2})$, $\alpha(\bvec{k})=J_x-\lambda \sin(k_y)$, and $\beta(\bvec{k})=J_y+\lambda \sin(k_x)$. Here, $A(\bvec{k})$ arises from a nearest-neighbor hopping term with hopping constant~$t$ that couples the orbital degrees of freedom, whereas $\alpha(\bvec{k})$ and $\beta(\bvec{k})$ contain the second-neighbor spin-orbit coupling contribution~$\lambda$. Further, $J_{\chi}=J_n (\bvec{n}\cdot\bvec{e}_{\chi})$ denotes the Cartesian components of the  antiferromagnetic vector $\bvec{J}=J_n \bvec{n}$ for a fixed direction of the N\'eel vector $\bvec{n}$. The Hamiltonian~\eqref{eq:hamiltonian} acts on a 4-dimensional space of electronic states with two degrees of freedom due to orbitals at the A and B sites and two spin degrees of freedom. The~$\sigma_{\chi}$ and~$\tau_{\chi}$ are Pauli matrices acting separately on the two spin and orbital degrees of freedom.  

The eigenstates of Hamiltonian~\eqref{eq:hamiltonian} are doubly degenerate with eigenenergies
\begin{equation}
     E(\bvec{k})=\pm \sqrt{A(\bvec{k})^2+\alpha(\bvec{k})^2+\beta(\bvec{k})^2+J_z^2},
     \label{eq:eigenenergies}
\end{equation}
which depend on $\bvec{k}$ through $A(\bvec{k})$, $\alpha(\bvec{k})$ and $\beta(\bvec{k})$. Following the procedure of Refs.~\cite{fabian_spin_1998,pientka_gauge_2012}, we fix the eigenstates in the two degenerate subspaces  by choosing them such that they diagonalize the spin operator in the direction of the N\'eel-vector at each $\bvec{k}$-point. More precisely, labeling the eigenstates at $\bvec{k}$ with band index $\nu$, we have $\langle\bvec{k}\nu|(\bvec{n}\cdot \boldsymbol{\sigma}) |\bvec{k}\nu'\rangle = 0$ for $\nu\neq\nu'$. Here, $\boldsymbol{\sigma}$ denotes the vector of Pauli matrices, which only acts on the spin degree of freedom. The resulting electronic Bloch states are therefore non-pure (or mixed) spin states of the form 
\begin{equation}
	|\bvec{k}\nu\rangle= a_{\nu}(\bvec{k}) |\sigma\rangle + b_{\nu}(\bvec{k})|\bar{\sigma}\rangle .
	\label{eq:a-b-coefficients}
\end{equation} If for one the doubly degenerate states at each $\bvec{k}$ the dominant spin component is~$\sigma=\uparrow$, then for the other one the dominant spin component is~$\sigma=\downarrow$. Conventionally, $a_{\nu}$ labels the large component and $b_\nu$ the small component, i.e., $|a_{\nu}(\bvec{k})| > |b_{\nu}(\bvec{k})|$.
 
The band structure~\eqref{eq:eigenenergies} is shown in Fig.~\ref{fig:band-structure}. We use a lattice constant of $a=0.45331 \, \mathrm{nm}$ which leads to a Brillouin zone with $k_{\mathrm{max}}=6.82\, \mathrm{nm}^{-1}$. We choose the hopping parameter $t= 100 \, \mathrm{meV}$, spin-orbit coupling parameter $\lambda=1.5 \times t$ and exchange coupling parameter $J_n=0.6 \times t$. The model exhibits two \emph{Dirac points} located at the border of the Brillouin zone at $\bvec{k}_1=(k_{\mathrm{max}}, 0.9\,\mathrm{nm}^{-1})$ and $\bvec{k}_2 = (k_{\mathrm{max}},6.0\,\mathrm{nm}^{-1})$. Only at these two points the upper and lower bands touch and the dispersion is approximately linear.  The parameter~$t$ has been reduced compared to Ref.~\cite{smejkal_electric_2017} in order to make the characteristic band structure feature less steep. The relative sizes of the spin-orbit and exchange coupling parameters with respect to $t$ are chosen close to the values of Ref.~\cite{smejkal_electric_2017}.

The spin expectation values are defined by $\langle \boldsymbol{\sigma}\rangle_{\bvec{k}\nu} = \langle\bvec{k}\nu|\boldsymbol{\sigma} |\bvec{k}\nu\rangle$
and shown for the upper band as arrows on top of the band structure in Fig.~\ref{fig:band-structure}. At each $\bvec{k}$, the spin expectation value of one of the degenerate bands is aligned anti-parallel to that of the other one. The whole band structure shows a non-trivial spin texture. For definiteness, we label the four bands as: ``upper''/``lower'' identifying the two sets of degenerate bands, and `$+$'/`$-$' according to the dominant spin character in each degenerate subspace. The label `$+$' indicates a spin direction that mostly points in the direction of the N\'{e}el vector and should not be confused with the sign of the eigenenergy~\eqref{eq:eigenenergies}. It should also be pointed out that the spin direction close to the border of the Brillouin zone may point in a different direction from the one used as the label.

\begin{figure*}[t!]
    \centering
    \includegraphics[width=\linewidth]{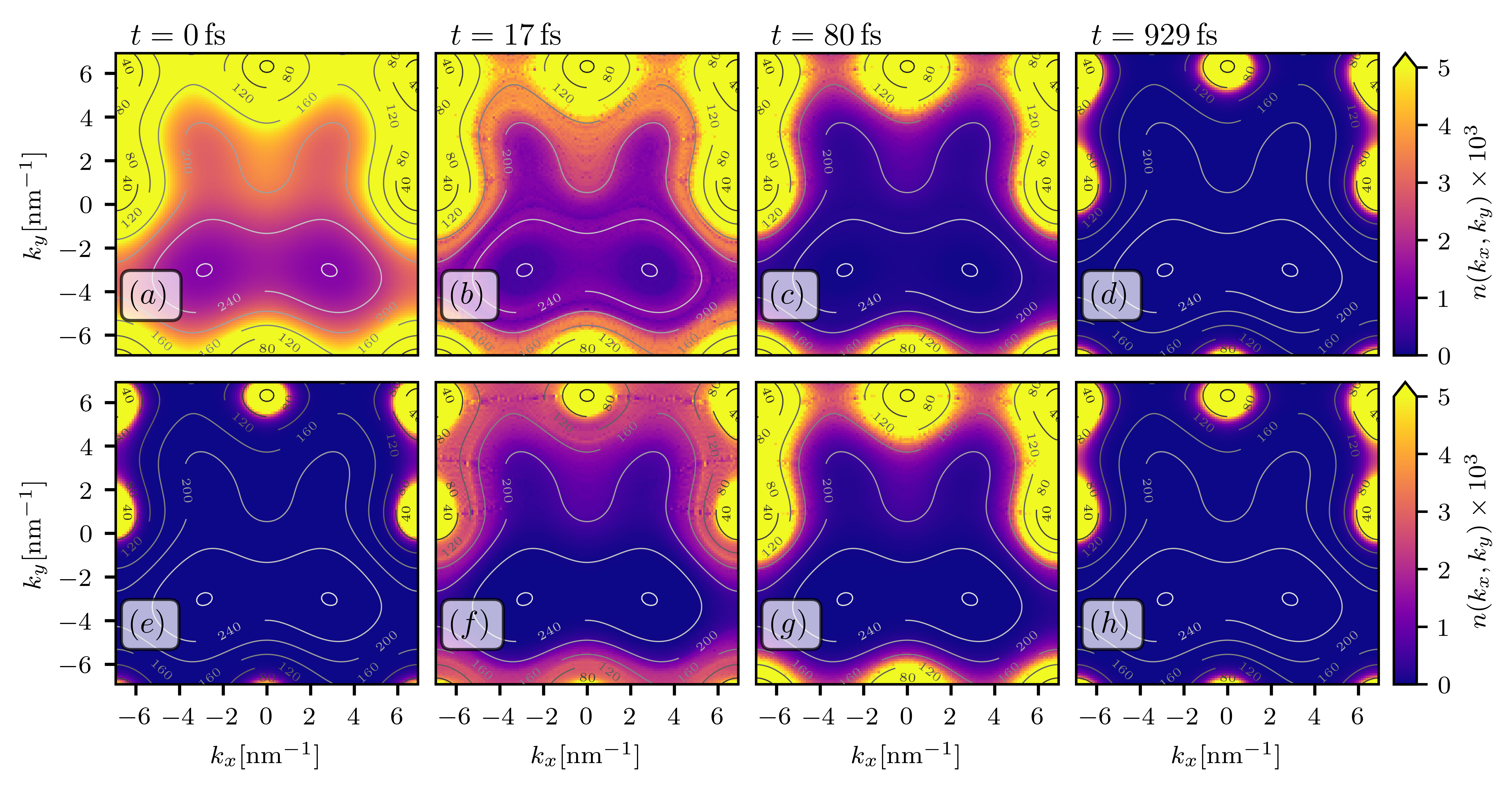}
    \caption{Color map of electron distributions $n_{\nu,\bvec{k}}$ vs.~$\bvec{k}$ for the upper degenerate bands (`$+$' band top row, `$-$' band bottom row) with a contour plot of the band structure as overlay. Contours and corresponding energy labels range from bright (280 meV) to dark lines (40 meV). Distributions are shown directly after the instantaneous excitation in the upper bands (a)\&(e), shortly after excitation (b)\&(f), when the upper bands are close to a quasiequilibrium (c)\&(g), and after relaxation within the upper bands (d)\&(h). 
    }
    \label{fig:k-resolved-electron-distributions}
\end{figure*}
For the incoherent electronic transitions after ultrafast optical excitation, there is usually a much larger scattering phase space available than for transport/dynamics close to the Fermi energy. The optically excited electrons sample the spin structure of the sequence of final states available to them. 
In order to describe these dynamics in the present paper, we consider electron scattering with longitudinal acoustic phonons. This scattering process can relax both carrier energy and momentum, and has been shown to play an important role on ultrafast timescales in ferromagnets~\cite{leckron_ultrafast_2017,mueller_feedback_2013,baral_re-examination_2016,essert_electron-phonon_2011,carva_ab_2011,koopmans_explaining_2010}. The fundamental dynamical quantities are the electronic distributions $n_{\bvec{k}}^{\nu} = \langle c_{\bvec{k},\nu}^{\dagger}c_{\bvec{k},\nu} \rangle$, i.e., the diagonal elements of the single-particle density matrix defined in terms of creation and annihilation operators  $c_{\bvec{k},\nu}^{(\dagger)}$ for electrons in the state $|\bvec{k}, \nu  \rangle $ with crystal momentum $\hbar \bvec{k}$ and band index~$\nu$.

The equation of motion (EOM) for the electronic distributions due to the interaction with phonons is used in the form 
\begin{widetext}
\begin{equation}
	\begin{split}
		\frac{\partial}{\partial t} n_{\bvec{k}}^{\nu} = 
		\frac{2\pi}{\hbar} \sum_{\bvec{k}' \nu'} \left|g_{\bvec{k}' \nu' , \bvec{k} \nu}\right|^2 \Big[ 
		&\big( ( 1+ N_{\bvec{k}' -\bvec{k}} ) (1 - n_{\bvec{k}}^{\nu}) n_{\bvec{k}'}^{\nu'}  - N_{\bvec{k}'-\bvec{k}} ( 1 -n_{\bvec{k}'}^{\nu'}) n_{\bvec{k}}^{\nu} \big)
		\delta (\Delta E_-\big|_{\bvec{k}'\bvec{k}}^{\nu'\nu}) \\
		- &\big(( 1+ N_{\bvec{k} -\bvec{k}'} ) ( 1 - n_{\bvec{k}'}^{\nu'} ) n_{\bvec{k}}^{\nu}  - N_{\bvec{k}-\bvec{k}'} ( 1 -n_{\bvec{k}}^{\nu} ) n_{\bvec{k}'}^{\nu'} \big)
	    \delta (\Delta E_+\big|_{\bvec{k}'\bvec{k}}^{\nu'\nu} )  \Big],
	\end{split}
	\label{eq:eom-evaluated}
\end{equation}
\end{widetext}
where $\Delta E_{\pm}$ is a shorthand notation for the energy differences $\Delta E _{\pm} = \varepsilon_{\bvec{k}}^{\nu} - \varepsilon_{\bvec{k}_1}^{\nu_1} \pm \hbar \omega_{q}$ that arise from energy conservation for the electron-phonon scattering process involving a phonon energy of $\hbar \omega_{q}$, where the phonon momentum is $q = |\bvec{k}_1-\bvec{k}|$. The phonons are assumed to form a bath so that their occupation numbers are always given by a Bose-Einstein distribution~$n_B(\hbar\omega_q)$ with an equilibrium temperature of $T^{(\mathrm{eq})} = 150$\,K. 
We compute the numerical solution of EOM~\eqref{eq:eom-evaluated} in the complete Brillouin zone to resolve the \emph{anisotropic} dynamics. For the phonon dispersion we assume a standard form $\hbar \omega_{\bvec{q}}= \hbar c_{\mathrm{pn}} \left|\sin\left(\frac{a}{2} q\right)\right|\frac{2}{a}$ with lattice constant $a$ and speed of sound $c_{\mathrm{pn}}=5 \, \mathrm{nm \; ps^{-1}}$.

For the electron-phonon matrix elements we use the expression
\begin{equation}
    g_{\bvec{k}\nu, \bvec{k}_1\nu_1} \simeq \sqrt{ \frac{\hbar}{2m_{\mathrm{ion}}\omega_{q}}} D q \langle \bvec{k}\nu | \bvec{k}_1\nu_1 \rangle,
    \label{eq:e-pn-matel}
\end{equation}
with  a deformation potential, $D=3$ eV~\cite{baral_re-examination_2016}. The overlap $\langle \bvec{k}\nu | \bvec{k}_1\nu_1 \rangle$ includes the dependence on the relative orientation of initial and final momenta of intra- and interband transitions.

The $\bvec{k}$ dependent spin structure of the model is determined by the spin-orbit coupling and exchange interaction contributions to the hamiltonian~\eqref{eq:hamiltonian}. It influences the change of the electronic ensemble spin that occurs due to incoherent scattering processes between Bloch states with different spin expectation values. This constitutes a dynamical version of the Elliott-Yafet spin-relaxation mechanism for paramagnetic metals and semiconductors, which, generally speaking, arises from the combination of electron-phonon scattering processes with spin-orbit coupling in the electronic states~\cite{Yafet1963,fabian_spin_1998,baral_re-examination_2016}. 

In order to provide a rough estimate of the spin changes possible due to this mechanism~\cite{fabian_spin_1998}, we note that the spin-relaxation time, which is only well defined \emph{close to equilibrium}, is proportional to the momentum scattering time and related to the spin-mixing coefficents~$b$ introduced in Eq.~\eqref{eq:a-b-coefficients}, which are larger for larger spin-orbit coupling~\cite{fabian_spin_1998}. In particular the relaxation time is approximately proportional to $|b|^2$, where $|b|^2$ denotes an average over the small component of the Bloch states involved in the dynamics. For our non-equilibrium case this means that the efficiency and therefore the time scale of spin-changing transitions is roughly controlled by the spin-orbit parameter~$\lambda$. 

\section{Results and Discussion}

\begin{figure}[b!]
    \centering
    \includegraphics[width=\linewidth]{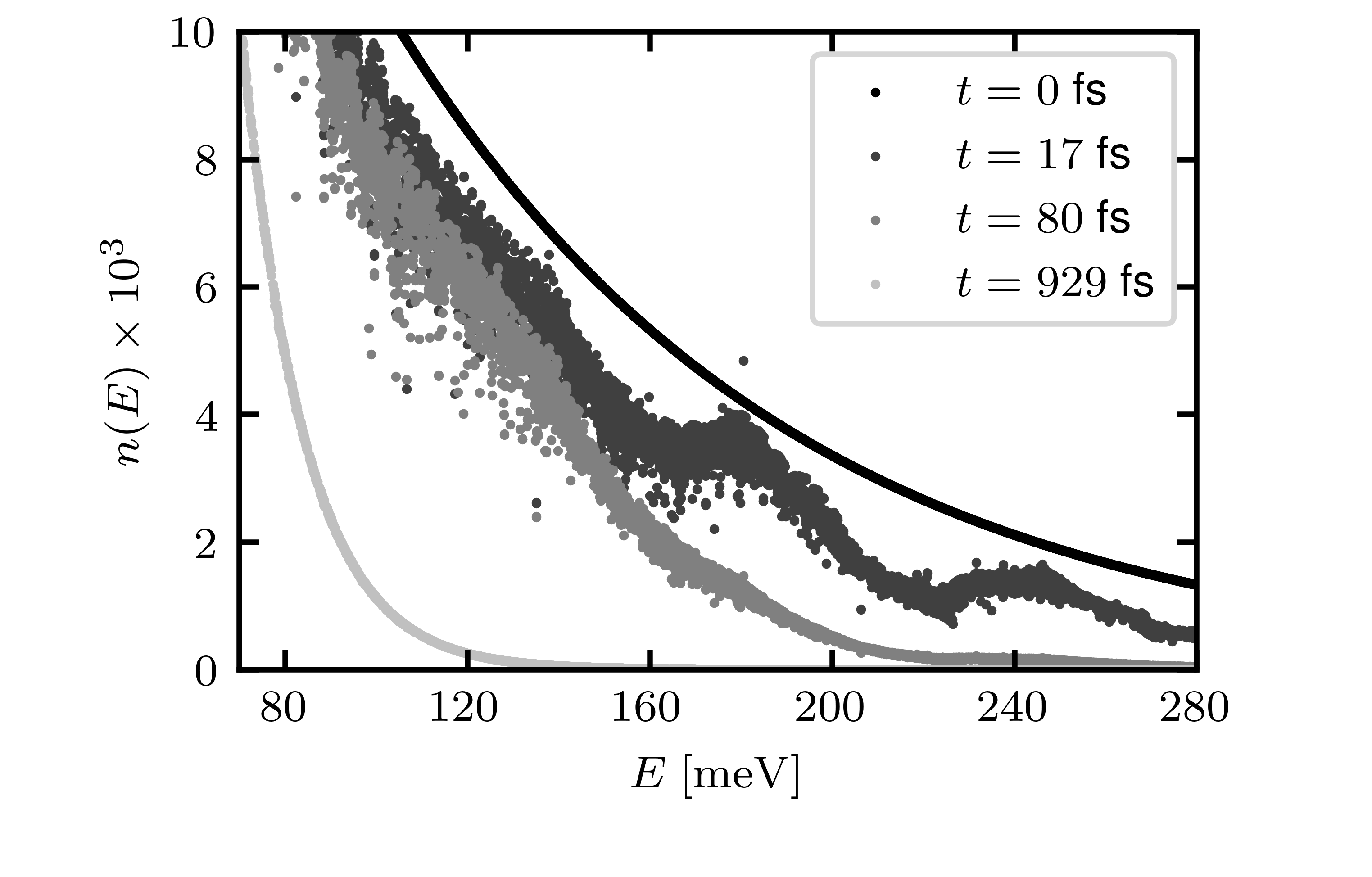}
    \caption{Electron distributions from the upper `$+$' band, as shown in Fig.~\ref{fig:k-resolved-electron-distributions} (a)--(d) plotted here vs. energy instead of~$\bvec{k}$. The contribution of each $\bvec{k}$-point is shown as a single dot. \emph{During} the scattering dynamics, electrons bunch up at certain energies and develop a pronounced anisotropy, which shows up here as a spread in the computed~$n(E)$ data.}
    \label{fig:e-resolved-electron-distributions}
\end{figure}

In order to study characteristics of spin dependent carrier scattering due to electron-phonon interactions we assume that the excitation process conserves the total ensemble spin expectation value, i.e., integrated over all bands. To this end, we will assume that electrons from the lower `$+$' band are excited to the upper `$+$', but that the memory of the detailed optical excitation process has been lost due to elastic electron-electron scattering and dephasing processes. We will thus approximate the excitation process as instantaneous, which is an assumption often made in similar calculations for ferromagnetic systems. There it has been shown that a very important characteristic of the excitation process is the energy per unit cell deposited by the pulse~\cite{essert_electron-phonon_2011,Stiehl_2022,Mrudul_Oppeneer_2024}. This important quantity is captured by assuming an instantaneous heating process for the study of scattering mechanisms. Here, we assume that it leads to the distributions shown for the upper `$+$' and `$-$' bands in Fig.~\ref{fig:k-resolved-electron-distributions} (a) and (e), respectively:  a cold ($T^{(\mathrm{eq})} = 150$\,K) Fermi-Dirac (FD) distribution in the unexcited `$-$' bands, see Fig.~\ref{fig:k-resolved-electron-distributions}(e), and a hot ($T^{(\mathrm{ex})} = 1500$\,K) FD distribution in the excited `$+$' bands, see Fig.~\ref{fig:k-resolved-electron-distributions}(a). These temperatures are determined from the kinetic energy of the electrons in the individual bands and characterize the different FD distributions. The initial carrier distributions thus depend only on the band energies. Starting from these initial distributions, the inelastic electron-phonon scattering dynamics are computed using Eq.~\eqref{eq:eom-evaluated}.
Note that the upper `$+$'-band contains additional electron density due to the excitation, leading to an equal but opposite change in electron density in the lower `$+$'-band, i.e., the creation of `holes'. We mainly discuss the dynamics for the upper bands in the following because the lower bands exhibit the same dynamics for holes.

Figure~\ref{fig:k-resolved-electron-distributions} shows snapshots of the electronic distribution functions in the 2D $k$-space as they evolve due to electron-phonon scattering. The panels (a)--(d) and (e)--(h) display the electron occupations for the upper `$+$' and `$-$' bands, respectively. Figs.~\ref{fig:k-resolved-electron-distributions} (a) and (e) show the initial state, i.e., the hot FD distribution in the `$+$' band (a) and cold FD in the `$-$' band (e). In Fig.~\ref{fig:k-resolved-electron-distributions}(e) the Dirac-points are visible at the left and right boundaries ($k_x = \pm k_{\mathrm{max}}$) because a small electron occupation in the upper bands can exist around these band crossings at finite temperatures. The region with comparatively high occupation $> 0.05$ in equilibrium around $k_x = 0\;\mathrm{nm}^{-1}$ and $k_y \approx 6\;\mathrm{nm}^{-1}$ is a \emph{local} minimum with a gap of 66\,meV between the upper and lower bands.

Figure~\ref{fig:k-resolved-electron-distributions}(b) reveals that electrons bunch up in the `$+$' band, as shown by the distinct belts of transient electron occupations located near the equipotential line around 240\,meV and between the equipotential lines for 160\,meV and 200\,meV that develop on a 10\,fs time scale and persist up to roughly 50 fs. Note that this is a striking feature of the dynamics as it occurs in regions of the band structure away from a local band minimum where one would expect an increase in density due to a lack of out-scattering phase space. The underlying process is electronic cooling, in which electrons scatter to lower energies within the `$+$' band, cf. Fig.~\ref{fig:k-resolved-electron-distributions}(b), as well as transitions from the `$+$' to the `$-$' band due to the finite overlap $\langle \bvec{k}+ | \bvec{k}_1- \rangle$ in the electron-phonon matrix element~\eqref{eq:e-pn-matel} as shown in Fig.~\ref{fig:k-resolved-electron-distributions}(f). In order to exhibit these features more clearly, we plot in Fig.~\ref{fig:e-resolved-electron-distributions} the electronic distributions in the `$+$'-band vs.\  energy instead of momentum $\bvec{k}$ for the same times as in Figs.~\ref{fig:k-resolved-electron-distributions} (a)--(d). Each dot represents the distribution at a $\bvec{k}$-point with the respective energy~\footnote{The outliers for $t = 17$ fs and $t = 80$ fs are numerical artifacts resulting from the finite $\bvec{k}$-grid.}

Figure~\ref{fig:e-resolved-electron-distributions} demonstrates two characteristics of electronic dynamics in such a band structure: First, the electrons bunch up in regions of $k$-space above the band minima, i.e., the Dirac-points and the local minimum at 33\,meV. Second, essentially momentum dependent distributions evolve from an initial state which depends only on the electronic energies, as indicated by the spread of the dots for $t=17$\,fs and $t=80$\,fs vs.\ the formation of a smooth curve for $t=0$\,fs. We have checked that the formation of these regions of transient non-equilibrium electron distributions is \emph{not} a numerical artifact, but a consequence of the available electronic scattering paths in this specific anisotropic band structure in combination with the phonon dispersion.  

\begin{figure}[t!]
    \centering
    \includegraphics{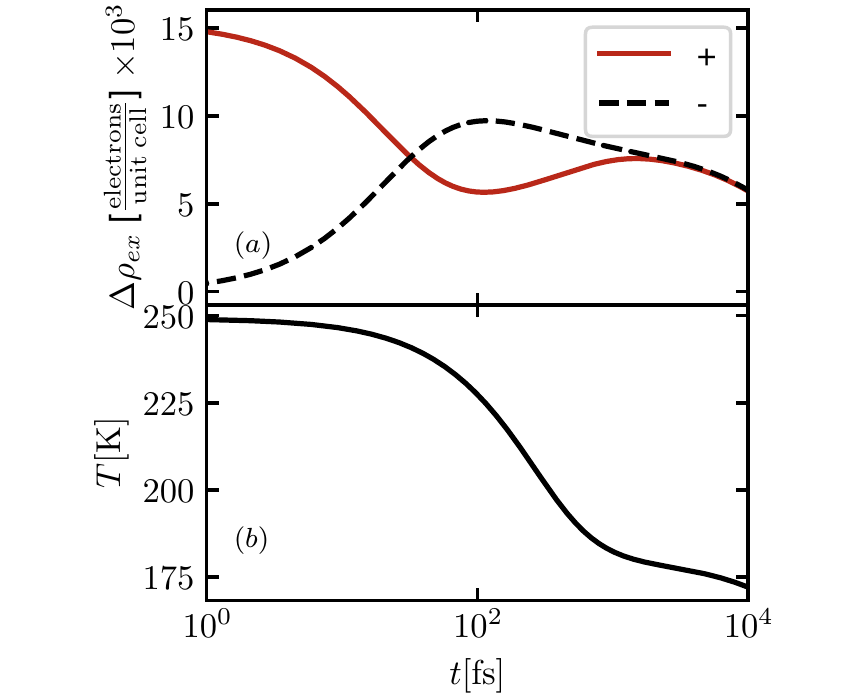}
    \caption{Time evolution of electron density in the upper bands (a) and temperature of the whole system (b).}
    \label{fig:density-temp}
\end{figure}
Returning to Fig.~\ref{fig:k-resolved-electron-distributions} we examine the remaining snapshots. At $t=80$\,fs, shown as Figs.~\ref{fig:k-resolved-electron-distributions}(c) and (g), a pronounced increase of the electron distribution in the upper `$-$' band and a decrease of the electron distribution in the upper `$+$' can be observed. For longer times, shown for $t=929$\,fs in Figs.~\ref{fig:k-resolved-electron-distributions}(d) and (h), the phonons continue to cool the distributions, until a quasi-equilibrium is reached; see also the discussion of Fig.~\ref{fig:density-temp} below. On a nanosecond time scale (not shown here) the additional electrons in the upper bands recombine with holes in the lower bands, and the whole system relaxes back to a complete equilibrium.

To obtain a qualitative characterization of the complicated $k$-dependent scattering dynamics we compute the carrier density and energy density in the upper bands. Figure~\ref{fig:density-temp}(a) shows the deviation of the electron density of the upper bands from its equilibrium value, $\Delta \rho_{\text{ex}}^\nu(t) = \rho_{\text{ex}}^\nu(t) - \rho_{\text{eq}}^\nu$. In Fig.~\ref{fig:density-temp}(b) we plot an effective temperature, which we determine by fitting a Fermi-Dirac (FD) distribution to the time-dependent electronic distribution functions by adjusting the temperature of the FD distribution so that the energy density of the electrons in both bands is reproduced. This effective temperature is intended as a measure of the energy density of the non-equilibrium distribution functions and not a temperature in the sense of thermodynamic equilibrium; it is also different from the temperatures of the electrons in the different bands quoted above, which are only used to characterize the excitation conditions. The integrated densities and energy densities contain contributions from all excited carriers, and are only weakly influenced by the dynamics at higher energies, which were analyzed in detail in Fig.~\ref{fig:e-resolved-electron-distributions}. Our excitation elevates roughly 0.015 electrons per unit cell from the lower `$+$' to the upper `$+$' band. This additional density in the excited upper `$+$'-band decreases as the density in the upper `$-$'-band increases which is caused by spin-flip scattering between the upper bands. This is essentially the signature of an Elliott-Yafet type spin flip or spin relaxation mechanism in this antiferromagnetic system and is mainly determined by the anisotropic band and spin structure of the single-particle Bloch states. The equilibration between the two upper degenerate bands in Fig.~\ref{fig:density-temp}(a) is accompanied by a fast decrease of the effective temperature in Fig.~\ref{fig:density-temp}(b) up to 2\,ps, which is a signature of pronounced change of the microscopic electronic distributions due to electron-phonon scattering processes. Afterwards the scattering to the lower bands leads to a slower decrease of the temperature in Fig.~\ref{fig:density-temp}(b).

\begin{figure}[t!]
    \centering
    \includegraphics[width=\linewidth]{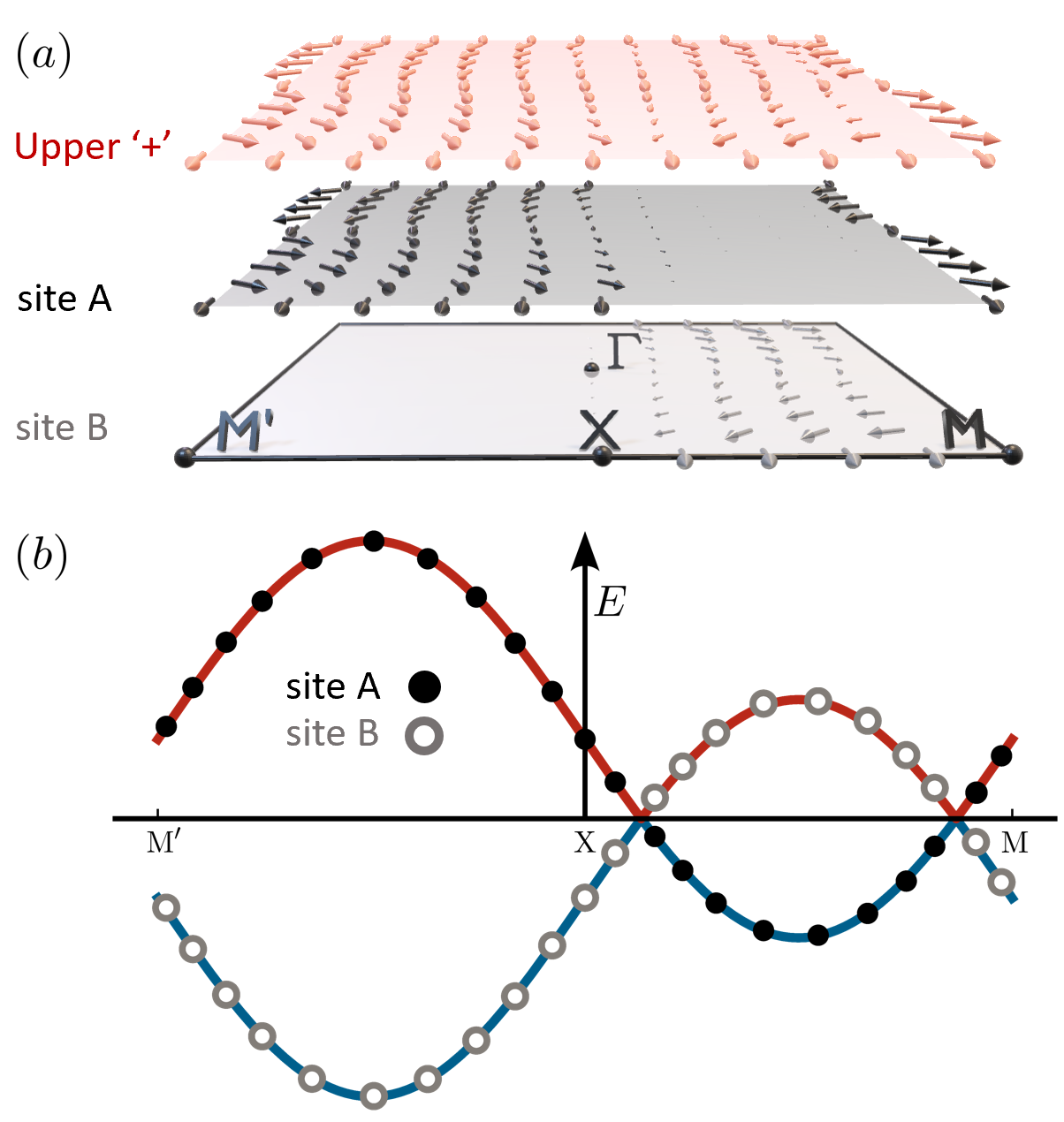}
    \caption{(a) Spin expectation values $\langle\boldsymbol{\sigma}\rangle_{\bvec{k}\nu}$ of the Bloch states for the upper `$+$'-band (red arrows), contributions from orbitals at site A (black arrows) and orbitals at site B (gray arrows).The Brillouin zone including relevant high symmetry points is shown as a square for site B . (b) Band structure and corresponding site-resolved spins along the high-symmetry line $M'\to X\to M$ for upper `$+$' band (red curve) and lower `$+$' band (blue curve). The black and grey dots correspond to the spin expectation value pointing out of the plane for contributions from site A and site B, respectively. .}
    \label{fig:site-resolved-spins}  
\end{figure}

So far we have investigated the features of the electronic dynamics in the band structure. 
We now turn to a characterization of the magnetic properties, for which the whole BZ including the Dirac points plays an important role. The antiferromagnetic spin polarization is a result of the staggered spins at A and B sites. From the dynamical electron distributions, we compute the $\bvec{k}$ dependent \emph{ensemble} spin expectation values, for which we also distinguish between contributions from A and B orbitals
\begin{equation}
    \begin{split}
       \Delta\langle  s_x \rangle_{A/B}(t) & = \sum_{\bvec{k},\nu} \langle \sigma_x \otimes P_{A/B} \rangle_{\bvec{k}\nu} \big[n_{\bvec{k}}^{\nu}(t)-n_{\bvec{k}}^{\nu}(-\infty)\big] \\
       &\equiv \langle s_x\rangle_{A/B} (t) - \langle s_x\rangle_{A/B}^{\mathrm{eq}}. 
   \end{split}
   \label{eq:delta-s-x}
\end{equation}
Here we used the projectors $P_{A/B}$  on orbitals at sites A and B, respectively. The top part of Fig.~\ref{fig:site-resolved-spins}(a) shows the spin orientation of the \emph{single-particle} spin~$\langle \boldsymbol{\sigma} \rangle_{\bvec{k}\nu}$ for the upper `$+$' band in the whole BZ. The remaining two graphs in Fig.~\ref{fig:site-resolved-spins}(a) illustrate the contributions from sites A and B, respectively. The A/B character of the band states changes along the lines connecting the Dirac points in the Brillouin zone.
\begin{figure}[t!]
    \centering
    \includegraphics[width=\linewidth]{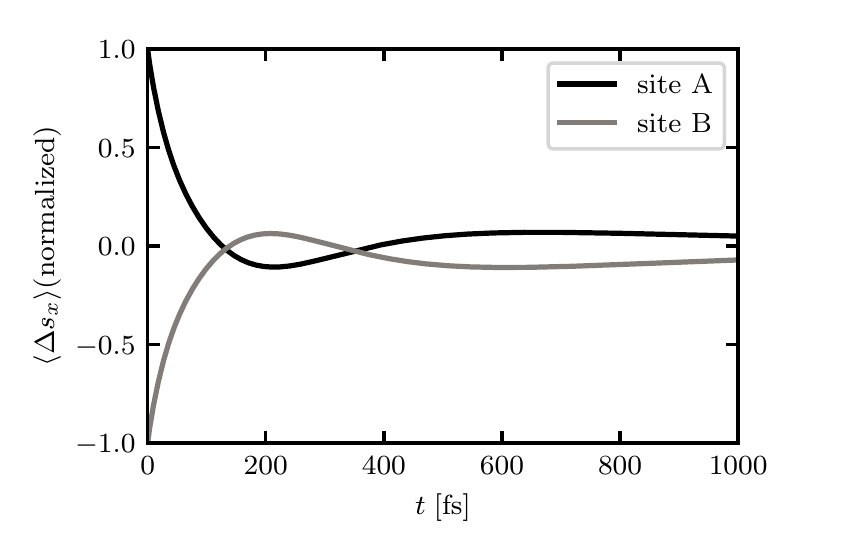}
    \caption{Change of ensemble spin expectation value in $x$ direction, projected on site A (black line) and site B (gray line), respectively. $\langle\Delta s_x\rangle_{A/B}(t)$ has been normalized to its $t=0$ value. }
    \label{fig:site-resolved}
\end{figure}

Figure~\ref{fig:site-resolved-spins}(b) connects the information contained in Fig.~\ref{fig:site-resolved-spins}(a) with the band structure for the $k$ space region around the Dirac point. Along the high symmetry lines $M'\to X \to M$ the dispersion of the upper and lower bands are shown together with the site projected spin expectation values and the corresponding orbital character for the `$+$' bands.
Figure~\ref{fig:site-resolved-spins} already indicates that the \emph{site-resolved} spin expectation value is changed when electrons from the lower `$+$' band are elevated to the upper `$+$' band by an excitation process that conserves the total spin. Such an excitation reduces the spin expectation value corresponding to site B and increases the one corresponding to site~A. This is a consequence of the nontrivial contributions of site A and site B to the bands in the antiferromagnetic Dirac structure visible in Fig.~\ref{fig:site-resolved-spins}(a). 

Figure~\ref{fig:site-resolved} plots the time evolution of the deviation of the site-resolved spin expectation value in the direction of the N\'eel vector ($x$), as defined in Eq.~\eqref{eq:delta-s-x}, normalized to its value at $t=0$. The dynamics of the site-resolved spin expectation values occur in opposite directions for A and B sites, so that there is an antiferromagnetic alignment at all times.   
The normalized $\langle \Delta s_x\rangle_A$, i.e., spin-change contribution from site A, goes through a dip at 200 fs, after which it increases again up to 700 fs. The relaxation back to the equilibrium value of 0 on the nanosecond time scale is associated with the \emph{inter}band equilibration. 
The change of the site-resolved spin occurs while the electron-phonon scattering redistributes electrons between the two upper bands, cf.~Fig.~\ref{fig:k-resolved-electron-distributions}.

We have normalized the site-resolved spin expectation value to its $t=0$ value in order to highlight the dynamics on the scale of the excitation. Normalizing to the equilibrium value $\langle s_x\rangle^{\text{eq}}_{A/B}$ would completely obscure this effect as $\langle s_x\rangle^{\text{eq}}_{A/B}$ is on the order of the carrier density, i.e., 2 electrons per unit cell, whereas the excited carrier density for our model excitation for the `$+$'-bands is approximately 0.015 electrons per unit cell.  

The change in the site-resolved spin expectation value of an antiferromagnet on ultrashort timescales is different from the case of a ferromagnet, where the band-resolved carrier distributions directly yield a contribution to the  magnetization dynamics on ultrashort timescales that can be detected in a straightforward fashion using the time-resolved magneto-optical Kerr effect~\cite{Beaurepaire.1996}.
In the antiferromagnet, such a direct relation between band-resolved charge dynamics in Fig.~\ref{fig:density-temp}(a) and spin dynamics~in Fig.~\ref{fig:site-resolved} does not exist. While the spin dynamics are not measurable directly by magneto-optical techniques as in ferromagnets, there exist experimental approaches that can probe them and thus make a comparison with our calculated results possible in principle. In addition to the well-known magnetic linear dichroism~\cite{Demsar-MLD-2021}, it has been shown that  AFM domains can be imaged using magneto-optical birefringence~\cite{AFM-optics-perspective}. This technique has been applied in pump-probe measurements on NiO also on ultrafast timescales~\cite{Meer-AFM,wust_indirect_unpublished_2022}, and it could also be applied to probe the dynamics investigated in the present manuscript, as it can track the change of the Néel vector, but not its absolute value.

\section{Conclusion and Outlook}

We have computed the electronic dynamics in an antiferromagnetic model system due to electron-phonon scattering after an instantaneous spin conserving excitation. The microscopic equations of motion for electron-phonon scattering were solved in the whole Brillouin zone including the pronounced anisotropy in the energy and spin structure that occurs in a Dirac semimetal antiferromagnet. As a result of its unique non-trivial band structure, transient anisotropic electron distributions can develop. We connected the antiferromagnetic spin dynamics via site-projected quantities with the electronic scattering dynamics on ultrashort timescales.
From the point of view of ultrafast magnetization dynamics, the results presented here establish a description of antiferromagnetic spin dynamics that is the counterpart to the ferromagnetic magnetization dynamics due to electron-phonon scattering in a fixed band structure~\cite{essert_electron-phonon_2011}, and the computed spin dynamics should be observable using present-day experimental set-ups. Our results are a first step towards the microscopic description of electron dynamics on ultrashort time scales in antiferromagnets. They can contribute to the interpretation of magneto-optical experiments on these systems. It is straightforward to extend the approach of this paper by considering \emph{ab-initio}  matrix elements and the optical excitation conditions but it is much more challenging to include electron-electron scattering and the interaction with magnetic excitations.

\begin{acknowledgments}
Funded by the Deutsche Forschungsgemeinschaft (DFG, German Research Foundation) 
– TRR 173 – 268565370 (projects A03, A08 and B03).
\end{acknowledgments}

%

\end{document}